\documentclass[11pt,preprint]{aastex}
\usepackage{epsfig,lscape}
\slugcomment{KSUPT-03/4 \hspace{0.5truecm} August 2003}

\newcommand{\lap}{\lower.5ex\hbox{$\; \buildrel < \over \sim \;$}}
\newcommand{\gap}{\lower.5ex\hbox{$\; \buildrel > \over \sim \;$}}
\tighten
 
\begin{document}
\title{Non-Gaussian Error Distribution of Hubble Constant Measurements}
\author{Gang Chen\altaffilmark{1}, J.~Richard Gott, III\altaffilmark{2}, 
        and Bharat Ratra\altaffilmark{1}}

\altaffiltext{1}{Department of Physics, Kansas State University, 116 Cardwell
                 Hall, Manhattan, KS 66506.}
\altaffiltext{2}{Princeton University Observatory, Peyton Hall, Princeton,
                 NJ 08544.}

\begin{abstract}
We construct the error distribution of Hubble constant ($H_0$) measurements
from Huchra's compilation of 461 measurements of $H_0$ and the WMAP
experiment central value $H_0$ = 71 km s$^{-1}$ Mpc$^{-1}$. This error
distribution is non-Gaussian, with significantly larger probability in
the tails of the distribution than predicted by a Gaussian distribution.
The 95.4 \% confidence limits are 7.0 $\sigma$ in terms of the quoted
errors. It is remarkably well described by either a widened $n = 2$ 
Student's $t$ distribution or a widened double exponential distribution.
These conclusions are unchanged if we use instead the central value 
$H_0$ = 67 km s$^{-1}$ Mpc$^{-1}$ found from a median statistics analysis 
of a major subset of $H_0$ measurements used here.
\end{abstract}
\keywords{cosmology: observation --- methods: statistical --- methods:
data analysis --- cosmology: distance scale --- large-scale structure 
of the universe}

\section{Introduction}

Measurements of the Hubble constant are a unique data set for statistical
analysis for two reasons. First, Huchra's compilation\footnote{
See cfa-www.harvard.edu/$\sim$huchra/.} 
with over 400 measurements is one of the largest collection of 
measurements of a single quantity. Second, the Hubble constant is now one 
of the more precisely determined cosmological parameters (see, e.g., 
Freedman et al.~2001; Bennett et al.~2003).

It is also of great interest to understand how well the Hubble constant has 
been measured, both because it is an important cosmological parameter 
and because of the role it plays in various cosmological tests, most 
importantly the expansion time test (see, e.g., Peebles \& Ratra 2003
for a review).

Assuming a value for the Hubble constant --- in the body of this paper we 
work with $H_0$ = 71 km s$^{-1}$ Mpc$^{-1}$, the central value from the 
combined WMAP and other data analysis of Spergel et al.~(2003)\footnote{ 
In the Appendix we summarize results from a similar analysis based on
the central value of $H_0$ = 67 km s$^{-1}$ Mpc$^{-1}$ from the Gott
et al.~(2001) median statistics analysis of a major, earlier subset of 
the $H_0$ measurements considered here, showing that our conclusions 
are not sensitive to the precise central value of $H_0$ assumed in the
estimated 10 \% (two standard deviation) range now under discussion 
(see, e.g., Gott et al.~2001).}
--- one may use Huchra's compilation of $H_{0i}{{+\sigma_i^{\rm u}} \atop
{-\sigma_i^{\rm l}}}$ (where $\sigma_i^{\rm u}$ and $\sigma_i^{\rm l}$
are the upper and lower one standard deviation error bars) to construct the 
distribution of errors of the Hubble constant measurements. This is a 
plot of the number of measurements as a function of the number of
standard deviations ($N_\sigma$) the measurement deviates from the actual
value $H_0$. Here
\begin{equation} 
  N_\sigma = {H_{0i} - H_0 \over \sigma_i^{\rm u}} ,
\end{equation}
when $H_{0i} < H_0$ and 
\begin{equation} 
  N_\sigma = {H_{0i} - H_0 \over \sigma_i^{\rm l}} ,
\end{equation}
when $H_{0i} > H_0$.

In our analysis here we use measurements from Huchra's compilation
up to and including measurement 2003.239. Deleting the four entries from 1924 
and 1925 that lack actual estimates of $H_0$, we use 461 published
estimates of $H_0$ in our analysis here, 40 \% more than the 331 used
in the analysis of Gott et al.~(2001). Observers often note that there could 
be unknown systematic errors, however authors' quoted errors have been used 
to evaluate the accuracy of $H_0$ estimates and so it is important to 
understand the quoted error distribution.

In $\S$ 2 we describe our analysis of this collection of 461 measurements, 
assuming that $H_0$ = 71 km s$^{-1}$ Mpc$^{-1}$, the central value from the 
combined WMAP and other data analysis of Bennett et al.~(2003) and Spergel et 
al.~(2003). For comparison, to show that the results are robust to small
changes in the true value of $H_0$, summary results from an analysis based 
on the central value of 
$H_0$ = 67 km s$^{-1}$ Mpc$^{-1}$ from the Gott et al.~(2001) median 
statistics study are presented in the Appendix. We conclude in $\S$ 3.

\section{Analysis}

Figure 1 shows the distribution of deviations of the 461 measurements
from the central WMAP value of $H_0$ = 71 km s$^{-1}$ Mpc$^{-1}$, in units
of the quoted standard deviation of the measurement. This is the error 
distribution 
of the $H_0$ measurements; the left panel shows the signed error distribution 
and the right panel shows the absolute magnitude of the errors (the
distribution in the right panel is symmetric about $|N_\sigma| = 0$).
These error distributions have significant tails: there are numerous
measurements 5 and even 10 standard deviations away. More precisely,
in the signed error distribution of Fig.~1$a$ 68.3 \% and 95.4 \% of the
probability lies in the range $-2.4 \leq N_\sigma \leq 1.1$ and 
$-7.0 \leq N_\sigma \leq 6.7$, respectively, and for the absolute magnitude
error distribution of Fig.~1$b$ the corresponding limits are $|N_\sigma| 
\leq 1.9$ and $|N_\sigma| \leq 7.0$, respectively. An alternative 
characterization of the tails of this distribution is provided by the 
fraction of data within the $|N_\sigma|$ = 1 and 2 ranges, which for the 
distribution shown in Fig.~1$b$ is 48 \% and 69 \%, respectively. These 
are impressively high (nearly half the observed values are within one 
standard deviation of 71 km s$^{-1}$ Mpc$^{-1}$) but still clearly at odds 
with what is expected for a Gaussian distribution.

It is of interest to quantify how well the data of Fig.~1 are fit by 
various simple distribution functions, and to determine the parameters of 
these functions that result in the best fit to the data. To do this we 
proceed as follows. For our purposes it is useful to maximize the number 
of data points in each bin as well as the number of bins. This is perhaps 
best accomplished by using 21 bins (close to the square root of 461), 
labelled by integer $j$ that runs from 1 to 21, and adjusting the 
widths of the bins, $\Delta|N_\sigma|_j$, to ensure equal expected 
probability (for the assumed distribution function) in each bin. Thus,
for an assumed distribution (such as a Gaussian) we construct 21 bins
such that the expected number of data points in each bin would be 21.95.
Then we compare with the number of data points observed in each of the 
21 bins and do a $\chi^2$ analysis, as discussed in the next paragraph. 
(Since the number expected in each bin is large compared to unity, 
a $\chi^2$ analysis is justified.) With this prescription, the data 
binning depends on the assumed probability distribution function, 
$P(|N_\sigma|)$ (in this paper we present results only from the fit to 
the symmetric absolute error distribution, e.g., that in Fig.~1$b$).

To estimate goodness of fit we use the assumed probability distribution 
function to compute the expected number of measurements in each bin $j$,
$N P(|N_\sigma|_j)$, where $N = 461$ is the total number of measurements.  
Since there are a finite number of measurements in each bin, they should
be Poisson
distributed with mean value $N P(|N_\sigma|_j)$ for the $j^{\rm th}$ bin.
For the Poisson distribution the variance $\sigma_j^2$ is equal to the mean 
hence the total $\chi^2$ is 
\begin{equation} 
  \chi^2 = \sum^{21}_{j = 1} {{ \left[M(|N_\sigma|_j) - N P(|N_\sigma|_j)
    \right]^2} \over N P(|N_\sigma|_j)} ,
\end{equation}
where $M(|N_\sigma|_j)$ is the observed number of measurements in each bin.
We shall tabulate the reduced $\chi^2$, $\chi^2_\nu = \chi^2/\nu$, where
$\nu$ is the number of degrees of freedom, i.e., the number of bins (21)
less the number of constraints and fitting parameters. Given $\chi^2_\nu$
and $\nu$ one may compute the probability that the assumed distribution 
well describes the spread of the measurements. In the computation of this 
probability we assume that the bins are uncorrelated, which is not 
necessarily true (since lower rungs of the distance ladder introduce 
correlations in subsets of the measurements). It is therefore wise to 
place quantitative emphasis on just the $\chi^2_\nu$ values and use the 
corresponding probabilities as simply a qualitative indicator of goodness 
of fit.  

We consider four probability distribution functions and as mentioned above
focus on the absolute magnitude error distribution, as in Fig.~1$b$, so all
distributions we consider will be centered at $|N_\sigma| = 0$. One constraint
that must be satisfied is that the total number of measurements must sum
to 461. Since we consider 21 bins and normalize to fit the total number of
measurements, a probability distribution function with no free parameters 
will have $\nu = 20$ degrees of freedom. 

Even though we have noted the existence of extended tails in the error
distributions of Fig.~1, it is natural --- perhaps pavlovian --- to first
consider the Gaussian distribution, initially with width chosen so that 
$|N_\sigma| = 1$ corresponds to one standard deviation, and then with a 
scale factor to vary the width of the distribution. That is, we take as
probability distribution function the Gaussian expression
\begin{equation} 
  P(|N_\sigma|) = {1 \over \sqrt{2\pi}} {\rm exp}\left[-|N_\sigma|^2/2\right] ,
\end{equation}
for the case where $|N_\sigma| = 1$ is equivalent to one standard deviation,
and then consider the function $P(|N_\sigma|/S)$ where $S$ is a scale factor
that is adjusted to minimize $\chi^2$. (We allow $S$ to vary over the range
0.5 to 3 in steps of 0.1 when computing $\chi^2$.) In the first case there
are no additional free parameters so $\nu = 20$; the scale factor $S$ is 
an additional free parameter in the second case so here we have $\nu = 19$ 
degrees of freedom. Figure 2 shows the measurement error histograms and
the best-fit Gaussians, both normalized to unit area. Numerical values are
listed in Table 1. These show that if $H_0$ = 71 km s$^{-1}$ Mpc$^{-1}$
then the measurement error distributions are extremely poorly fit by
a Gaussian, even if the Gaussian width is allowed to be a free parameter.
Interestingly, if the width is allowed to float while minimizing $\chi^2$
it favors 1.8, i.e., the assumed distribution favors identifying $|N_\sigma|
= 1.8$ with one standard deviation, almost double the value one would 
naively infer from the measurement errors, thus perhaps indicating that 
in this Gaussian case it might not be unreasonable to roughly double the 
quoted error bars, consistent with our earlier discussion of extended 
tails. This situation might profitably be contrasted with what happened in the 
early days of cosmic microwave background spatial anisotropy measurements,
where a number of models fit the measurements extremely well, perhaps 
indicating that the error bars had been over estimated (Ganga, Ratra,
\& Sugiyama 1996). In any case, it is very unlikely that the 
$H_0$ measurement errors are described by a Gaussian distribution. (Note that 
the probability is a little higher for the $H_0$ = 67 km s$^{-1}$ Mpc$^{-1}$
case, but even here a Gaussian distribution is a very poor fit.)  

The fact that the error distribution of Hubble constant measurements is 
non-Gaussian does not necessarily imply an underlying non-Gaussianity in
the measurement errors. Rather, the distribution tells us something about 
the observers ability to correctly estimate systematic and statistical 
uncertainties.  

Figure 2 indicates that the distribution of Hubble constant measurement 
errors has a more extended tail than is predicted by a Gaussian probability
distribution. Perhaps the most well-known distribution with an extended tail
is the Cauchy, or Lorentzian, or Breit-Wigner distribution,
\begin{equation} 
  P(|N_\sigma|) = {1 \over \pi} {1 \over 1 + |N_\sigma|^2} ;
\end{equation}
we also consider the case $P(|N_\sigma|/S)$ where the scale factor
$S$ is allowed to vary while $\chi^2$ is minimized. Figure 3 shows the 
data and best-fit Cauchy distributions, and numerical values are listed in
Table 1. Unlike the Gaussian case, the Cauchy distribution can not be 
rejected; it is acceptable at 9.9 \% or 8.7 \% depending on whether 
$S$ is fixed to unity or allowed to vary (and it does significantly better
at $H_0$ = 67 km s$^{-1}$ Mpc$^{-1}$). However, it is clear from Fig. 3
that the Cauchy distribution has greater probability in the extended tails
than does the Hubble constant measurements error distribution. The Cauchy
distribution has a similar central peak, with a 50 \% chance that 
$|N_\sigma| < 1$, but a 95.4 \% chance that $|N_\sigma| < 13.8$ instead
of 7.0 as observed. It would therefore be beneficial to search for a 
distribution that has broader tails than the Gaussian one but narrower 
than the Cauchy case.

A Cauchy distribution with $S = 1$ would result if the errors were Gaussian 
distributed, observers took measurements free of systematic errors, divide 
their data into two parts, used each half to produce two independent 
estimates of the Hubble constant, $H_1$ and $H_2$, and produced a mean
estimate $H_{\rm m} = (H_1 + H_2)/2$ with an error estimate (standard 
deviation of the mean) of $\sigma_{\rm m} = |H_1 - H_2|/2$. If $H_1$
and $H_2$ are drawn from an underlying Gaussian distribution centered on 
the true value $H_{\rm t}$ then $(H_{\rm m} - H_{\rm t})/\sigma_{\rm m}$
is distributed like a Cauchy distribution with $S=1$. That gives a 50 \% 
chance that $H_{\rm m}$ is within 1 $\sigma_{\rm m}$ of the true value and a 
95 \% chance that $H_{\rm m}$ is within 12.7 $\sigma_{\rm m}$ of the true 
value. The large tails result because in a Gaussian distribution there is 
an appreciable chance that  $|H_1 - H_2|/2$ will be significantly less than 
the true sigma for the distribution. In this scenario the observer is really
using the self-consistency of her observations to set the error bars. If
one measures the distance to two galaxies using cepheids, and gets two 
values for the Hubble constant that are close to each other, one may well
be tempted to think that one's method has the high degree of accuracy 
implied by the observed value of $\sigma_{\rm m} = |H_1 - H_2|/2$. Indeed,
if one estimated the errors by other means (estimated uncertainties in
measuring the observed quantities required to measure the Hubble constant,
along with standard propagation of errors) and one got an error significantly
larger than $|H_1 - H_2|/2$ then one might be suspicious that one should be
so lucky as to obtain such a small value of $|H_1 - H_2|/2$. Yet rarely,
such lucky coincidences do occur and it is precisely these cases that cause 
the large tails in the Cauchy distribution. The Cauchy distribution with
$S = 1$ is acceptable at 9.9 \%, but not a good fit and there are other 
distributions that are better fits. As we have noted, a Cauchy 
distribution with $S = 1$ would result from a true Gaussian distribution 
if the observer divided his data into two parts, used the data itself to set error bars, and made the mistake of assuming the errors should be distributed 
according to a Gaussian distribution rather than the Student $t$ distribution
(which for the case of two data points is the $n = 1$ Student $t$ 
distribution, or the Cauchy distribution). This prompts us to investigate 
the general Student $t$ distributions. 

Student's $t$ distribution is
\begin{equation} 
  P_n(|N_\sigma|) = {\Gamma\left[(n+1)/2\right] \over\sqrt{\pi n}\, 
  \Gamma(n/2)} 
  {1 \over \left(1 + |N_\sigma|^2/n\right)^{(n+1)/2}} ,
\end{equation}
where $n$ is positive and $\Gamma$ the Gamma function. We also consider the 
distribution $P_n(|N_\sigma|/S)$ where the scale factor $S$ is allowed to 
vary while $\chi^2$ is minimized. When $n \rightarrow \infty$ Student's $t$
distribution becomes the Gaussian distribution and for $n = 1$ it is the Cauchy
distribution. For $1 < n < \infty$ Student's $t$ distribution has narrower 
tails than the Cauchy case but broader ones than the Gaussian distribution, 
just as wanted. We have fit Student's $t$ distribution to the $H_0$ 
measurement errors data while allowing $n$ to take on integer values between 
2 and 6 (and sometimes going up to 30), so in this case we have one additional 
parameter and hence one less degree of freedom. We find $n=2$ always
minimizes the value of $\chi^2$ and so show this case in Fig.~4 and Table 
1. From Table 1 we see that if the scale factor $S$ is held at unity Student's
$t$ distribution is an unlikely fit to the data, especially if 
$H_0$ = 71 km s$^{-1}$ Mpc$^{-1}$. However, if $S$ is allowed to vary as 
$\chi^2$ is minimized, Student's $t$ distribution with $n=2$ is an excellent 
fit to the $H_0$ measurements error distribution, and Fig.~4$b$ shows that 
there is very good agreement between the expected and measured counts in the 
last bin.

The final probability density distribution we consider is the double 
exponential or Laplace distribution,
\begin{equation} 
  P(|N_\sigma|) = {1 \over 2} e^{-|N_\sigma|} .
\end{equation}
This falls off less rapidly than the Gaussian distribution but faster 
than the Cauchy distribution. The sample median is the best estimator 
for the mean of this distribution (Eadie et al. 1971). The results of 
the fit are shown in Fig.~5 and listed in Table 1. As in the case for 
Student's $t$ distribution with $n=2$, when $S$ is held fixed at unity 
the double exponential is an unacceptable fit to
the $H_0$ measurements error distribution but when $S$ is allowed to vary
it is an excellent fit to the data.  

In the first paragraph of this section we noted that in Fig.~1$b$ 
68.3 \% and 95.4 \% of the probability lies in the range $|N_\sigma| 
\leq 1.9$ and $|N_\sigma| \leq 7.0$, respectively, and the $|N_\sigma| 
\leq 1$ and $|N_\sigma| \leq 2$ ranges include 48 \% and 69 \% of the
data points, respectively. (See the Appendix for the corresponding 
numbers for the $H_0$ = 67 km s$^{-1}$ Mpc$^{-1}$ case.) Tables 2 
and 3 show the related limits for the various probability density 
distributions we consider in this paper. These numerical values provide 
another indication of the non-Gaussianity of the Hubble constant 
measurement error distribution.

\section{Conclusion}

Our analysis of a perhaps unique (because of its size) data set, the 
measurement errors of all available estimates of the Hubble constant,
makes for some interesting conclusions. If all observers have done 
perfect jobs at estimating their errors and the true errors were Gaussian, 
as might be expected, then the distributions in Fig.~1 should be 
Gaussian with standard deviation of unity.

First, and perhaps not totally unexpectedly, the errors in 
the Hubble constant are not Gaussianly distributed, even if the scale
factor $S$ is allowed to vary when minimizing $\chi^2$. At the minimum
value of $\chi^2$, $S \sim 2$, suggesting it might be reasonable to 
roughly double the magnitude of $H_0$ measurement error bars. Early
observers using inferior equipment or techniques would have larger
errors, but knowing that their methods were uncertain should have 
established larger error bars. As methods improved the measurements 
become more accurate but the stated error bars become smaller. Early or late 
observers are at no relative disadvantage relative to others. Indeed,
each observer has freedom to state her error bars and has a priori an
equal chance of having the true value occur within one standard deviation
of their result. Over-optimism would produce error bars that were too
small while over-conservatism would produce error bars that were too large.
Which occurs in practice? The results here suggest that astronomers were 
over-optimistic by almost a factor of 2. Why? In some case there were 
systematic errors of which the observers were simply unaware (such as 
mistaking HII regions for bright stars). In other cases, standard candles
were not as standard as imagined, leaving some steps in the distance 
ladder wrong by more than people thought. Also, using self consistency
in the data as a check on the errors can lead to large tails because it
occasionally induces one to be over-optimistic (the Student $t$ effect).
And the real data may have non-Gaussian tails (say in the luminosity 
of standard candles). In general over-conservatism (the urge to be right)
always competes with over-optimism (the urge to have more interesting 
limits). In the case of the Hubble constant astronomers were over-optimistic.
In a history-of-science context, it might be of interest to more closely 
examine the most deviant measurements of Fig.~1, those that have 
$|N_\sigma|$ larger than say 7, to understand why these are so deviant,
but this is not our purpose here.

The Hubble constant measurement history suggests that to be really 
sure (95.4~\%) you have to go to 7 $\sigma$. This may explain why 
some people are cautious upon hearing of a three standard deviation
result. It's not that they believe the errors but want to be more sure 
than 99.7~\%. It's that they suspect there is a large chance ($\sim$
50~\%) that the error bars may have been underestimated by a factor of 2 
or 3 and the chance it is really correct is consequently really 
significantly less than 99.7~\%. 
 
Second, an $n=2$ Student's $t$ distribution, with $S \sim 1.2 - 1.3$, or 
a double exponential distribution, with $S \sim 1.5 - 1.6$, are excellent fits 
to the $H_0$ measurement errors distribution, with $H_0$ = 67 km s$^{-1}$ 
Mpc$^{-1}$ having a somewhat higher probability than $H_0$ = 71 km s$^{-1}$ 
Mpc$^{-1}$.   

The Hubble constant measurement history gives an interesting example 
where we can access how trustworthy quoted errors might be in 
fundamental measurements. It would be interesting to study comparative 
examples from other fields. In particular, it would be interesting to 
know whether the $n=2$ Student $t$ distribution or the double 
exponential distribution also provides a good description of the 
measurement errors of other quantities.

\acknowledgments

We are grateful to J.~Huchra for the compilation of $H_0$ measurements 
and acknowledge useful discussions with A.~Kosowsky. GC and BR acknowledge 
support from NSF CAREER grant AST-9875031 and DOE EPSCoR grant 
DE-FG02-00ER45824. JRG acknowledges support from NSF grant AST-9900772.

\appendix
\section{$H_0$ = 67 km s$^{-1}$ Mpc$^{-1}$}

In the main body of the paper we assumed $H_0$ = 71 km s$^{-1}$ Mpc$^{-1}$,
the central value from the WMAP analysis. The WMAP $H_0$ error bars are
${ }{{+4} \atop {-3}}$ km s$^{-1}$ Mpc$^{-1}$ (Bennett et al.~2003).
$H_0$ is pinned down to only about
10 \% at two standard deviations (Gott et al.~2001) so it is reasonable
to find out how our conclusions depend on the value of $H_0$. In this
Appendix we use $H_0$ = 67 km s$^{-1}$ Mpc$^{-1}$, the central value from
the Gott et al.~(2001) median statistics analysis of a subset (331 
measurements prior to mid 1999) of the 461 measurements used here.\footnote{
The additional 130 measurements (an increase of 40 \%) shift the 
median value to $H_0$ = 68 km s$^{-1}$ Mpc$^{-1}$ (using all 461 
measurements); the small shift in the median after a 40 \% increase 
in the number of measurements considered is great tribute to its 
robustness. See Gott et al.~(2001), Podariu et al.~(2001), Avelino,
Martins, \& Pinto (2002), and Chen \& Ratra (2003) for other 
cosmological applications of median statistics.}
Figure 6 shows the $H_0$ measurements error distribution for the 
case when $H_0$ = 67 km s$^{-1}$ Mpc$^{-1}$. This distribution has a 
somewhat less prominent central peak than the $H_0$ = 71 km s$^{-1}$ 
Mpc$^{-1}$ case. In the signed error distribution of Fig.~6$a$ 68.3 \% 
and 95.4 \% of the probability lies in the range $-1.8 \leq N_\sigma \leq 
1.7$ and $-5.7 \leq N_\sigma \leq 7.9$, respectively, while for the 
absolute magnitude error distribution of Fig.~6$b$ the corresponding 
limits are $|N_\sigma| \leq 1.7$ and $|N_\sigma| \leq 7.5$, respectively. 
In Fig.~6$b$ the $|N_\sigma| \leq 1$ and $|N_\sigma| \leq 2$ ranges
include 51 \% and 72 \% of the data points, respectively. Again, these 
are at odds with what is expected for a Gaussian distribution.

Table 1 also lists the numerical fitting results for the $H_0$ = 67 km 
s$^{-1}$ Mpc$^{-1}$ case. As in the case when $H_0$ = 71 km s$^{-1}$ 
Mpc$^{-1}$, here again the $n=2$ Student's $t$ distribution and the double 
exponential distribution 
provide excellent fits to the $H_0$ error histogram when the scale
factor $S$ is allowed to vary when minimizing $\chi^2$. These two distributions
are shown in Figs.~7 and 8. It might be significant that the $H_0$ = 67 km 
s$^{-1}$ Mpc$^{-1}$ case always has a lower $\chi^2$ than the $H_0$ = 71 km s$^{-1}$ Mpc$^{-1}$ case, indicating perhaps that the median statistics
value determined from a large fraction of the data is more robust --- time
will tell. In any case, a comparison of the entries in Table 1 shows that 
the results presented here are robust with respect to small changes in the 
value of $H_0$.

\clearpage

\begin{deluxetable}{lccccccccccccc}
\tablecolumns{11} 
\tablewidth{0pt}
\tablecaption{Goodness of Fit Numerical Values}
\tablehead{
\colhead{} & \colhead{} & \multicolumn{4}{c}{$H_0$ = 71 km s$^{-1}$ Mpc$^{-1}$} & \colhead{} & \multicolumn{4}{c}{$H_0$ = 67 km s$^{-1}$ Mpc$^{-1}$} \\
\cline{3-6} \cline{8-11} \\
\colhead{function} & \colhead{} & \colhead{scale\tablenotemark{a}} &
\colhead{$\chi^2_\nu$\tablenotemark{b}} & \colhead{$\nu$\tablenotemark{b}} &
\colhead{prob. (\%)\tablenotemark{c}} & \colhead{} & \colhead{scale\tablenotemark{a}} &
\colhead{$\chi^2_\nu$\tablenotemark{b}} & \colhead{$\nu$\tablenotemark{b}} &
\colhead{prob. (\%)\tablenotemark{c}} \\
\colhead{(1)} & \colhead{} & \colhead{(2)} & \colhead{(3)} & 
\colhead{(4)} & \colhead{(5)} & \colhead{} & \colhead{(6)} & \colhead{(7)} & 
\colhead{(8)} & \colhead{(9)}}
\startdata
Gaussian & & 1   & 19.8 & 20 & $< 0.1$ & & 1   & 15.0 & 20 & $< 0.1$ \\
Gaussian & & 1.8 & 2.63 & 19 & $< 0.1$ & & 1.7 & 1.92 & 19 &  0.94 \\
Cauchy   & & 1   & 1.42 & 20 &   9.9 & & 1   & 1.10 & 20 &    35 \\
Cauchy   & & 1.1 & 1.46 & 19 &   8.7 & & 1.0 & 1.15 & 19 &    29 \\
$n=2$ Student's $t$ & & 1   & 2.58 & 19 & $< 0.1$ & & 1   & 1.56 & 19 & 5.7 \\
$n=2$ Student's $t$ & & 1.3 & 0.717 & 18 & 80 & & 1.2 & 0.326 & 18 &  99.7 \\
Double Exponential & & 1   & 7.12 & 20 &  $< 0.1$ & & 1   & 5.11 & 20 & $< 0.1$ \\
Double Exponential & & 1.5 & 0.501 & 19 &   96 & & 1.6 & 0.325 & 19 &  99.7 
\enddata
\tablenotetext{a}{Scale (factor) $S = 1$ corresponds to the case when 
$|N_\sigma| = 1$ corresponds to one standard deviation for a Gaussian
distribution, otherwise the width of the distribution is allowed
to vary with the scale factor as $\chi^2$ is minimized.}
\tablenotetext{b}{$\chi^2_\nu$ is the $\chi^2$ per degree of freedom,
where $\nu$ is the number of degrees of freedom.}
\tablenotetext{c}{Probability that a random sample of data points
drawn from the assumed distribution yields a value of $\chi^2_\nu$ 
greater than or equal to the observed value for $\nu$ degrees of 
freedom. The computation assumes that the bins are uncorrelated, which
is not necessarily true, so the probabilities should not be taken at 
face value but merely as qualitative indicatiors of goodness of fit.}
\end{deluxetable}

\clearpage

\begin{deluxetable}{lcccccccc}
\tablecolumns{9} 
\tablewidth{0pt}
\tablecaption{$|N_\sigma|$ Limits}
\tablehead{
\colhead{} & \colhead{} & \multicolumn{3}{c}{$H_0$ = 71 km s$^{-1}$ Mpc$^{-1}$} & \colhead{} & \multicolumn{3}{c}{$H_0$ = 67 km s$^{-1}$ Mpc$^{-1}$} \\
\cline{3-5} \cline{7-9} \\
\colhead{function} & \colhead{} & \colhead{scale\tablenotemark{a}} &
\colhead{68.3 \%\tablenotemark{b}} & \colhead{95.4 \%\tablenotemark{b}} &
\colhead{} & \colhead{scale\tablenotemark{a}} &
\colhead{68.3 \%\tablenotemark{b}} & \colhead{95.4 \%\tablenotemark{b}} \\
\colhead{(1)} & \colhead{} & \colhead{(2)} & \colhead{(3)} & 
\colhead{(4)} & \colhead{} & \colhead{(5)} & \colhead{(6)} & 
\colhead{(7)}}
\startdata
Gaussian & & 1   & 1.0 & 2.0 & & 1   & 1.0 & 2.0 \\
Gaussian & & 1.8 & 1.8 & 3.6 & & 1.7 & 1.7 & 3.4 \\
Cauchy   & & 1   & 1.8 & 14  & & 1   & 1.8 & 14  \\
Cauchy   & & 1.1 & 2.0 & 15  & & 1.0 & 1.8 & 14  \\
$n=2$ Student's $t$ & & 1   & 1.3 & 4.5 & & 1   & 1.3 & 4.5 \\
$n=2$ Student's $t$ & & 1.3 & 1.7 & 5.9 & & 1.2 & 1.6 & 5.4 \\
Double Exponential  & & 1   & 1.2 & 3.1 & & 1   & 1.2 & 3.1 \\
Double Exponential  & & 1.5 & 1.7 & 4.6 & & 1.6 & 1.8 & 4.9 \\
\tableline
Observed            & &     & 1.9 & 7.0 & &     & 1.7 & 7.5
\enddata
\tablenotetext{a}{Scale (factor) $S = 1$ corresponds to the case when 
$|N_\sigma| = 1$ corresponds to one standard deviation for a Gaussian
distribution, otherwise the width of the distribution is allowed to 
vary with the scale factor as $\chi^2$ is minimized. For a given
distribution the derived limits depend only on the value of $S$.}
\tablenotetext{b}{Upper $|N_\sigma|$ limits that include 68.3 \% and
95.4 \% of the probability.}
\end{deluxetable}

\clearpage

\begin{deluxetable}{lcccccccc}
\tablecolumns{9} 
\tablewidth{0pt}
\tablecaption{Expected Fraction of Data Points with $|N_\sigma| \leq 1$ and $|N_\sigma| \leq 2$}
\tablehead{
\colhead{} & \colhead{} & \multicolumn{3}{c}{$H_0$ = 71 km s$^{-1}$ Mpc$^{-1}$} & \colhead{} & \multicolumn{3}{c}{$H_0$ = 67 km s$^{-1}$ Mpc$^{-1}$} \\
\cline{3-5} \cline{7-9} \\
\colhead{function} & \colhead{} & \colhead{scale\tablenotemark{a}} &
\colhead{$|N_\sigma| \leq 1$\tablenotemark{b}} & \colhead{$|N_\sigma| \leq 2$\tablenotemark{b}} &
\colhead{} & \colhead{scale\tablenotemark{a}} &
\colhead{$|N_\sigma| \leq 1$\tablenotemark{b}} & \colhead{$|N_\sigma| \leq 2$\tablenotemark{b}} \\
\colhead{(1)} & \colhead{} & \colhead{(2)} & \colhead{(3)} & 
\colhead{(4)} & \colhead{} & \colhead{(5)} & \colhead{(6)} & 
\colhead{(7)}}
\startdata
Gaussian & & 1   & 0.68 & 0.95 & & 1   & 0.68 & 0.95 \\
Gaussian & & 1.8 & 0.42 & 0.73 & & 1.7 & 0.44 & 0.76 \\
Cauchy   & & 1   & 0.50 & 0.71 & & 1   & 0.50 & 0.71 \\
Cauchy   & & 1.1 & 0.47 & 0.68 & & 1.0 & 0.50 & 0.71 \\
$n=2$ Student's $t$ & & 1   & 0.58 & 0.82 & & 1   & 0.58 & 0.82 \\
$n=2$ Student's $t$ & & 1.3 & 0.48 & 0.74 & & 1.2 & 0.51 & 0.76 \\
Double Exponential  & & 1   & 0.63 & 0.87 & & 1   & 0.63 & 0.87 \\
Double Exponential  & & 1.5 & 0.49 & 0.74 & & 1.6 & 0.47 & 0.71 \\
\tableline
Observed            & &     & 0.48 & 0.69 & &     & 0.51 & 0.72
\enddata
\tablenotetext{a}{Scale (factor) $S = 1$ corresponds to the case when 
$|N_\sigma| = 1$ corresponds to one standard deviation for a Gaussian
distribution, otherwise the width of the distribution is allowed to 
vary with the scale factor as $\chi^2$ is minimized. For a given
distribution the derived limits depend only on the value of $S$.}
\tablenotetext{b}{Fraction of data points with $|N_\sigma| \leq 1$ or
$|N_\sigma| \leq 2$.}
\end{deluxetable}

\begin{figure}
\centerline{\epsfig{file=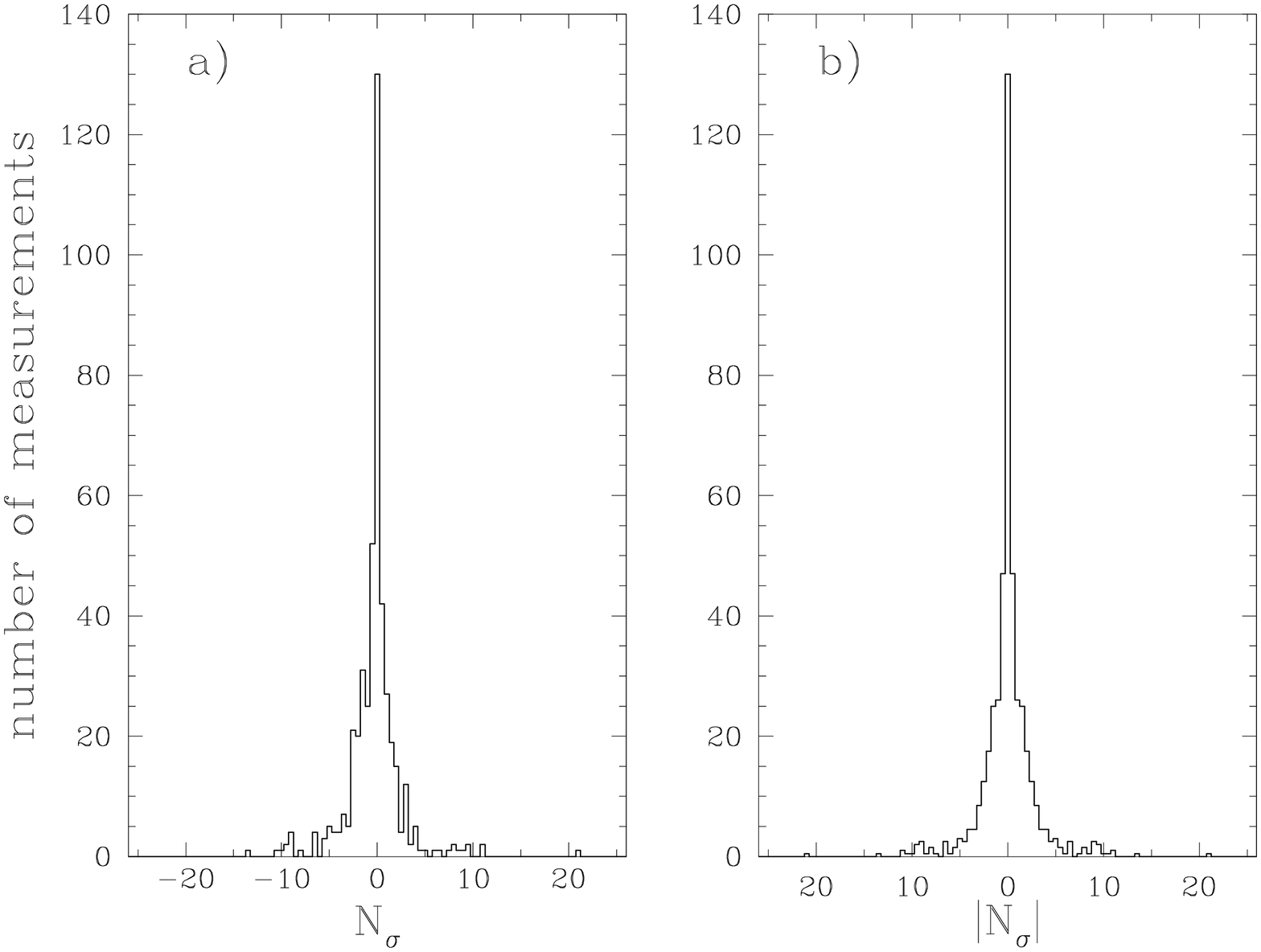,width=15.5cm,angle=270}}
\caption{Number of measurements (in half standard deviation bins) away from 
the central value of $H_0$ = 71 km s$^{-1}$ Mpc$^{-1}$ estimated by the WMAP
collaboration. Left panel $a)$ shows the sign of the deviation while right
panel $b)$ shows only the magnitude of the deviation. In panel $a)$ bins
with positive (negative) $N_\sigma$ correspond to measurements where $H_0$
is measured to be higher (lower) than 71 km s$^{-1}$ Mpc$^{-1}$. }
\label{f1}
\end{figure} 

\begin{figure}
\centerline{\epsfig{file=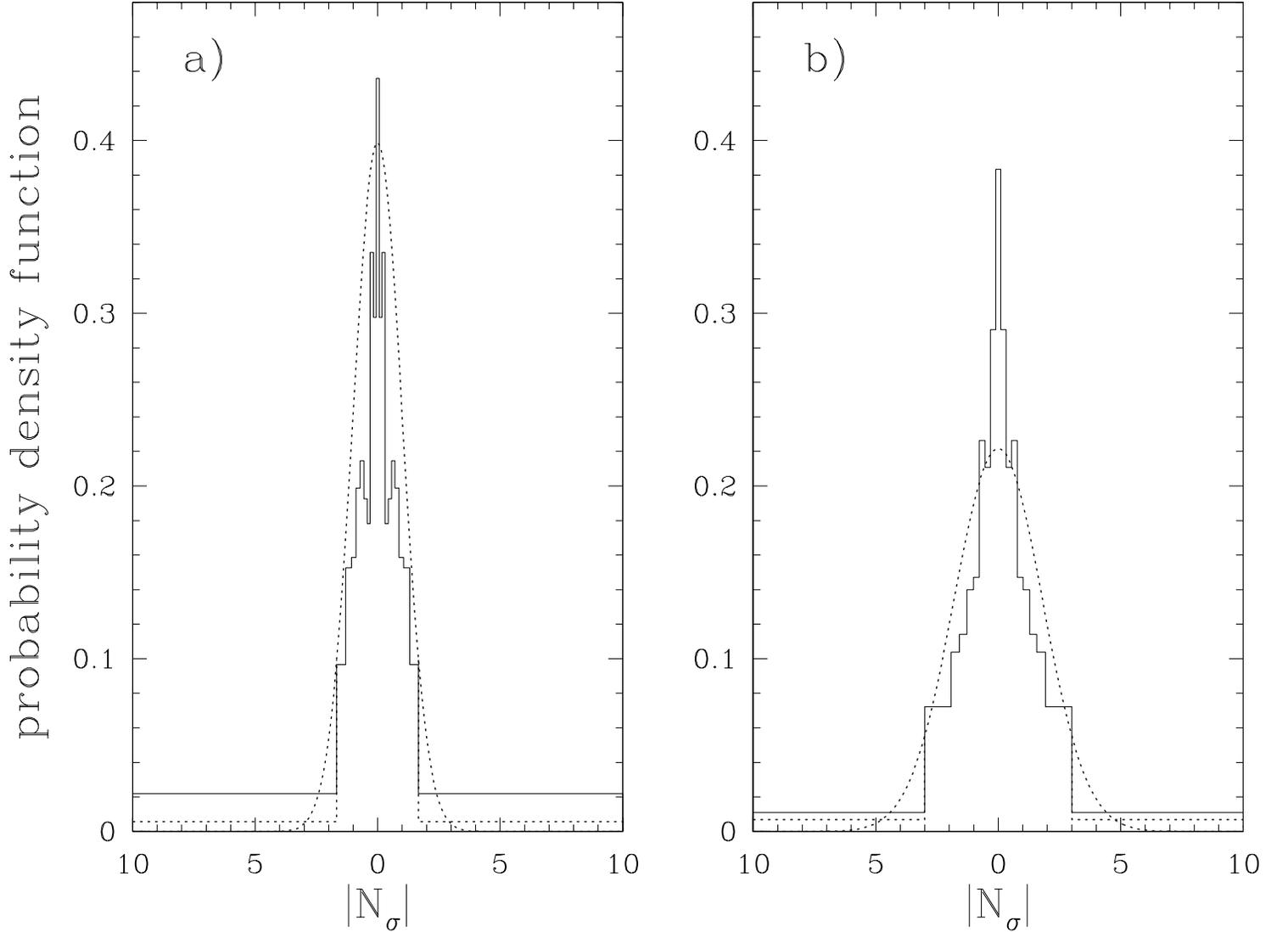,width=15.5cm,angle=270}}
\caption{Binned data (solid lines) and best-fit Gaussian probability
distribution functions (dotted lines) for $H_0$ = 71 km s$^{-1}$ Mpc$^{-1}$ 
estimated by the WMAP collaboration, all normalized to unit area. The
binning and therefore the data histogram shapes depend on the assumed 
probability distribution function
(see text). Left panel $a)$ shows a Gaussian distribution with scale factor
$S=1$ such that $|N_\sigma| = 1$ corresponds to one standard deviation; right
panel $b)$ allows $S$ to vary as $\chi^2$ is minimized and the best fit
value of $S = 2.0$ is shown. For ease
of visualization, the extreme ends of the left- and right-most bins (solid
lines) have been brought in from $|N_\sigma| = \infty$ to $|N_\sigma| = 10$,
with the heights adjusted to ensure that the probabilities in the bins are
unchanged. The dotted horizontal and vertical lines demarcate the predicted
probability for these last bins for the assumed Gaussian distribution.} 
\label{f2}
\end{figure}

\begin{figure}
\centerline{\epsfig{file=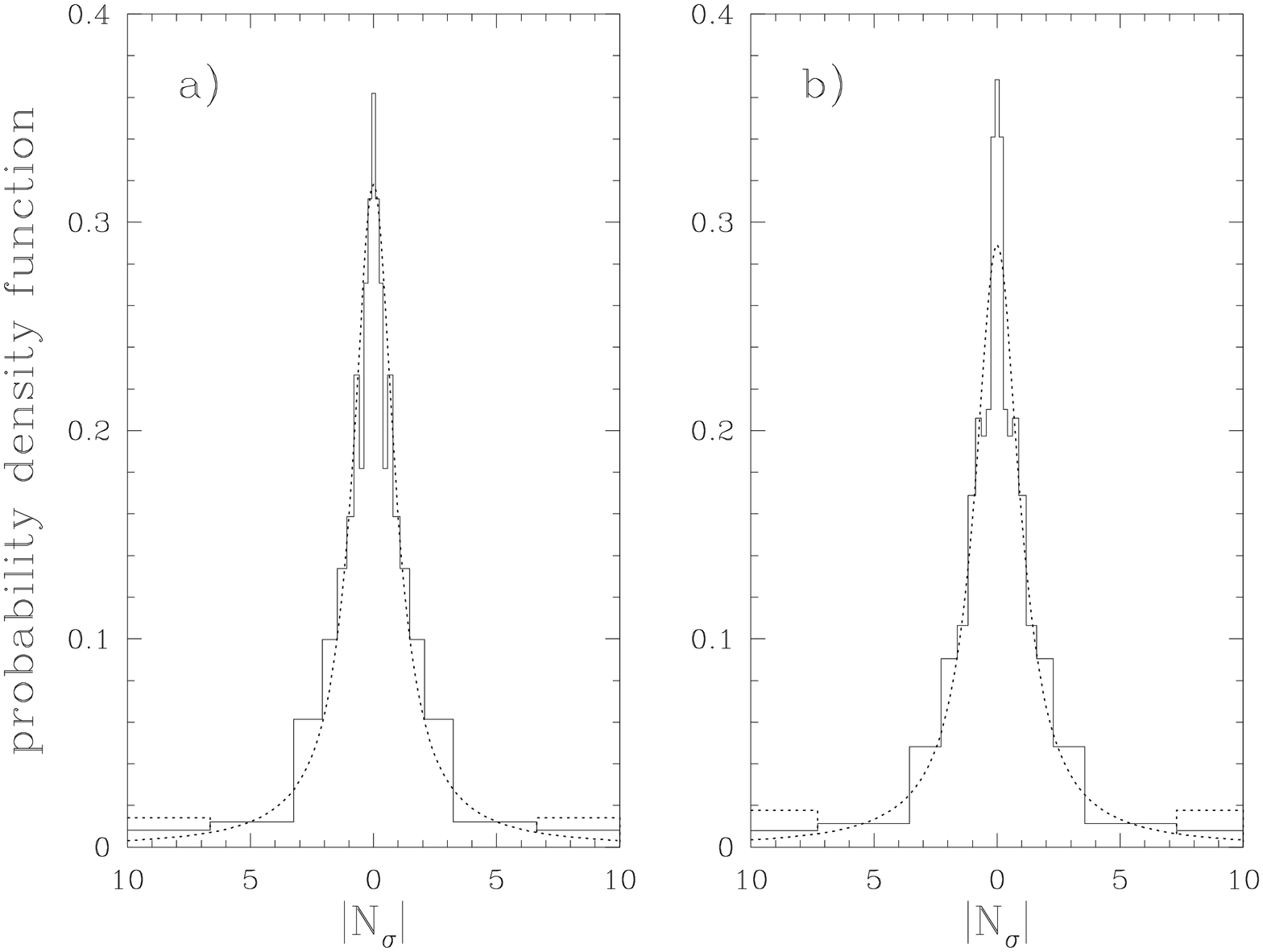,width=15.5cm,angle=270}}
\caption{Binned data (solid lines) and best-fit Cauchy probability
distribution functions (dotted lines) for $H_0$ = 71 km s$^{-1}$ Mpc$^{-1}$ 
estimated by the WMAP collaboration, all normalized to unit area. See Fig.~2 
caption for more details. Left panel $a)$ shows a Cauchy distribution with 
scale factor $S=1$; right panel $b)$ allows $S$ to vary as $\chi^2$ is 
minimized and the best fit value of $S = 1.1$ is shown. The dotted 
horizontal and vertical lines demarcate the predicted
probability for the last bins for the assumed Cauchy distribution.}
\label{f3}
\end{figure}

\begin{figure}
\centerline{\epsfig{file=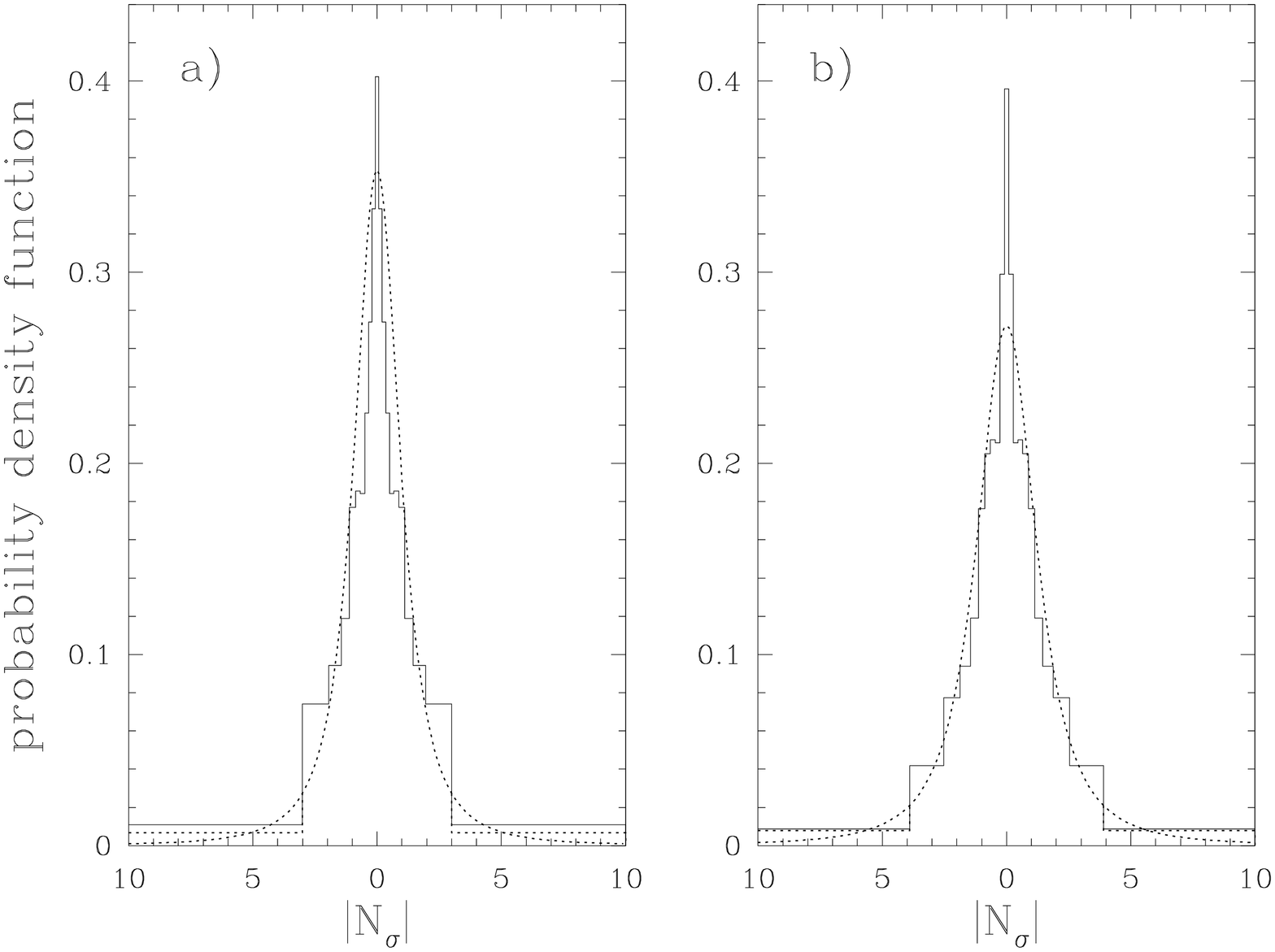,width=15.5cm,angle=270}}
\caption{Binned data (solid lines) and best-fit $n=2$ Student's $t$ probability
distribution functions (dotted lines) for $H_0$ = 71 km s$^{-1}$ Mpc$^{-1}$ 
estimated by the WMAP collaboration, all normalized to unit area. See Fig.~2 
caption for more details. Left panel $a)$ shows a distribution with 
scale factor $S=1$; right panel $b)$ allows $S$ to vary as $\chi^2$ is 
minimized and the best fit value of $S = 1.3$ is shown. The dotted horizontal 
and vertical lines demarcate the predicted
probability for the last bins for the assumed distribution.}
\label{f4}
\end{figure}

\begin{figure}
\centerline{\epsfig{file=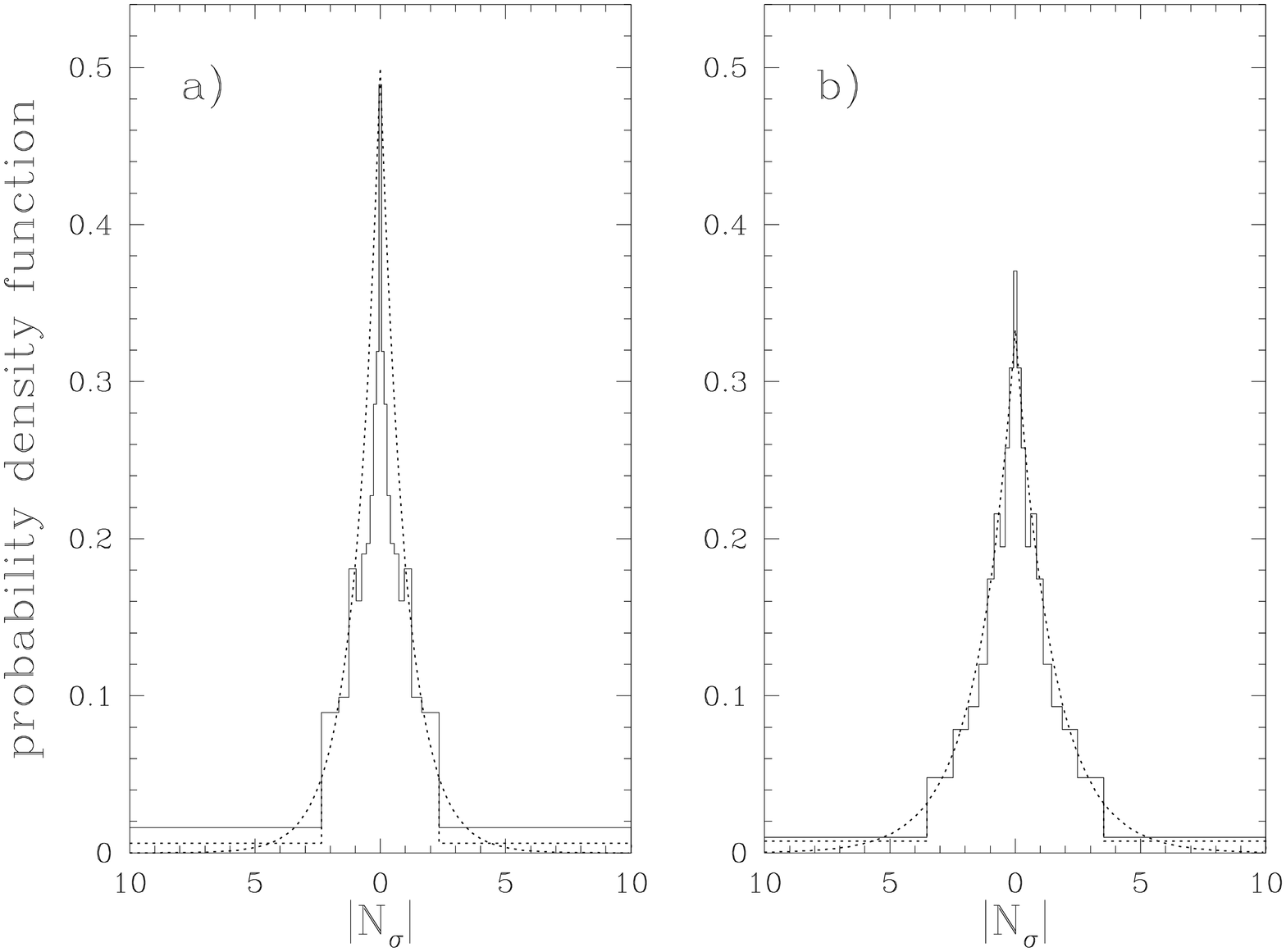,width=15.5cm,angle=270}}
\caption{Binned data (solid lines) and best-fit double exponential probability
distribution functions (dotted lines) for $H_0$ = 71 km s$^{-1}$ Mpc$^{-1}$ 
estimated by the WMAP collaboration, all normalized to unit area. See Fig.~2 
caption for more details. Left panel $a)$ shows a distribution with 
scale factor $S=1$; right panel $b)$ allows $S$ to vary as $\chi^2$ is 
minimized and the best fit value of $S = 1.5$ is shown. The dotted 
horizontal and vertical lines demarcate the predicted
probability for the last bins for the assumed double exponential distribution.}
\label{f5}
\end{figure}

\begin{figure}
\centerline{\epsfig{file=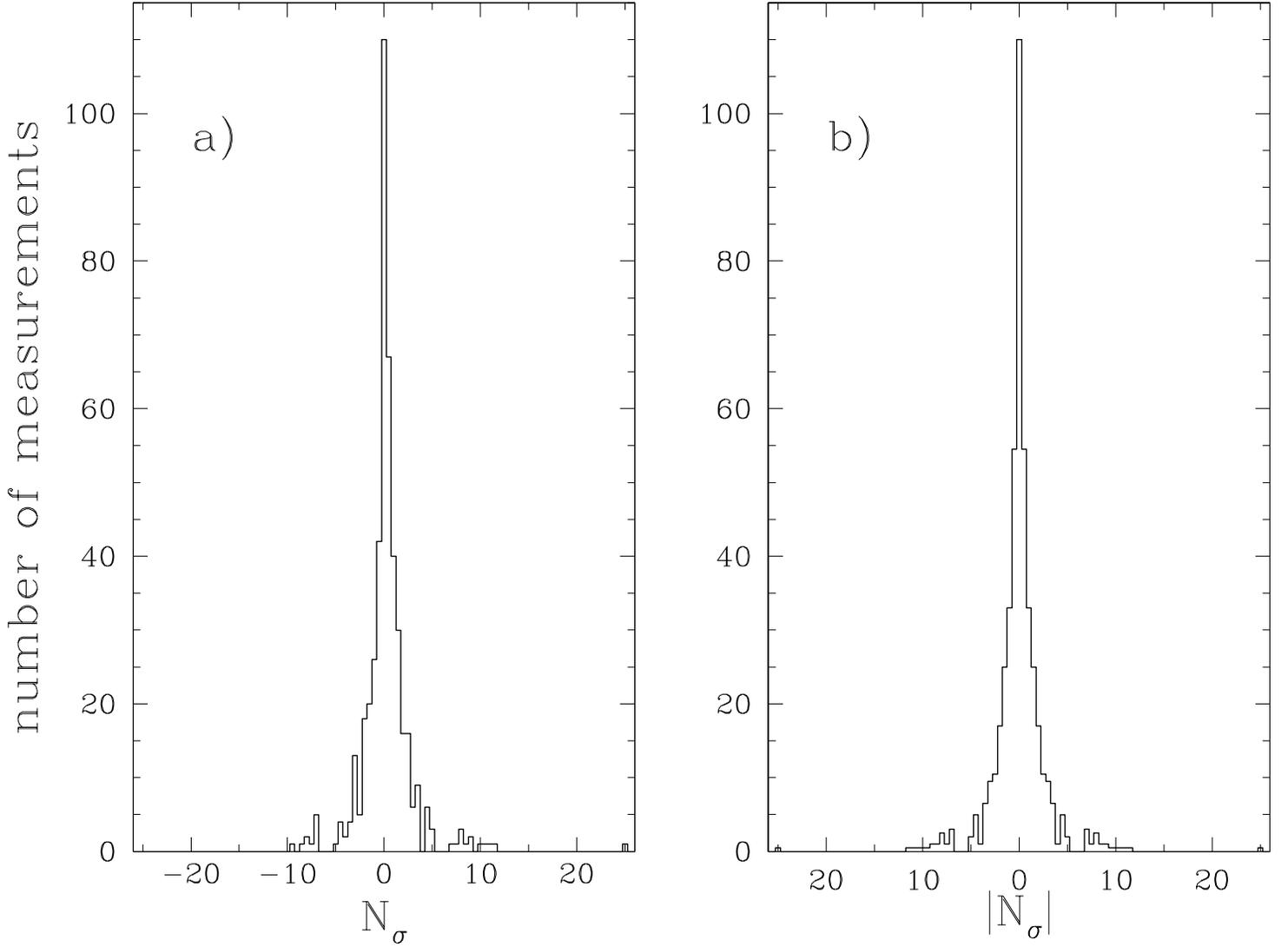,width=15.5cm,angle=270}}
\caption{Number of measurements (in half standard deviation bins) away from 
the central value of $H_0$ = 67 km s$^{-1}$ Mpc$^{-1}$ estimated using median
statistics on a major subset of the 461 measurements used in this paper.
Left panel $a)$ shows the sign of the deviation while right
panel $b)$ shows only the magnitude of the deviation.}
\label{f6}
\end{figure}

\begin{figure}
\centerline{\epsfig{file=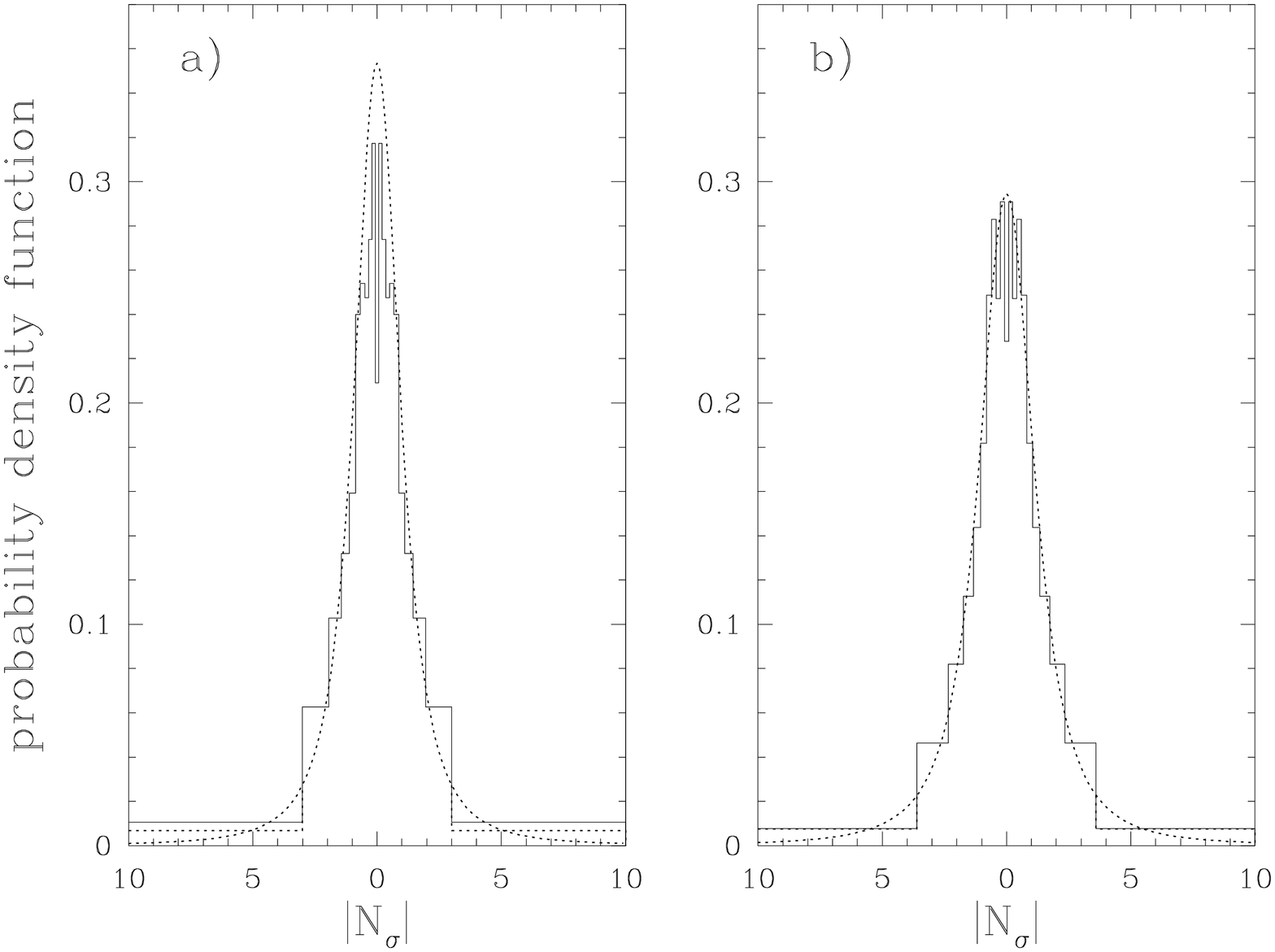,width=15.5cm,angle=270}}
\caption{Binned data (solid lines) and best-fit $n=2$ Student's $t$ probability
distribution functions (dotted lines) for $H_0$ = 67 km s$^{-1}$ Mpc$^{-1}$
estimated using median statistics, all normalized to unit area. See Fig.~2 
caption for more details. Left panel $a)$ shows a distribution with 
scale factor $S=1$; right panel $b)$ allows $S$ to vary as $\chi^2$ is 
minimized and the best fit value of $S = 1.2$ is shown. The dotted horizontal 
and vertical lines demarcate the predicted
probability for the last bins for the assumed distribution.}
\label{f7}
\end{figure}

\begin{figure}
\centerline{\epsfig{file=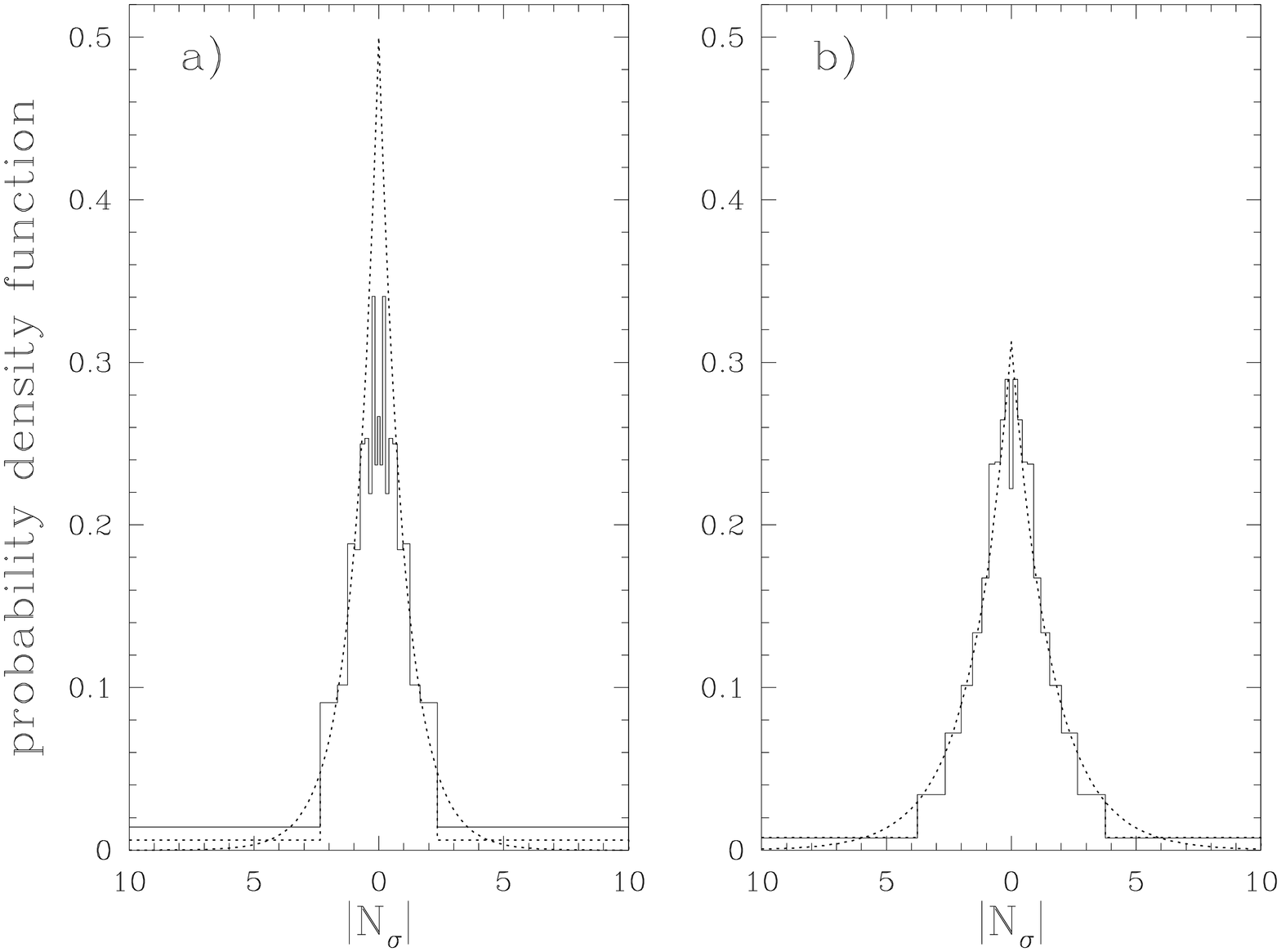,width=15.5cm,angle=270}}
\caption{Binned data (solid lines) and best-fit double exponential probability
distribution functions (dotted lines) for $H_0$ = 67 km s$^{-1}$ Mpc$^{-1}$
estimated using median statistics, all normalized to unit area. See Fig.~2 
caption for more details. Left panel $a)$ shows a distribution with 
scale factor $S=1$; right panel $b)$ allows $S$ to vary as $\chi^2$ is 
minimized and the best fit value of $S = 1.5$ is shown. The dotted horizontal 
and vertical lines demarcate the predicted
probability for the last bins for the assumed distribution.}
\label{f8}
\end{figure}

\end{document}